\documentclass{appolb}
\usepackage{epsfig}
\begin{document}
\title{On Gamov states of   $\Sigma^+$ hyperons}
\author{S{\l}awomir Wycech
\address{National Centre  for Nuclear Studies, Ho\.za 69, 00-681 Warsaw,
Poland}
\\{Kristian  Piscicchia}
\address{CENTRO FERMI - Museo Storico della Fisica e Centro Studi e Ricerche "Enrico Fermi", Roma, Italy\\
INFN Laboratori Nazionali di Frascati, Frascati (Roma), Italy\\ }}
\maketitle

\begin{abstract}
Both the FINUDA and AMADEUS experiments evidenced a low $\Sigma^+$ momentum component when investigating $\Sigma^+ \pi^-$ pairs produced in $K^-$ nuclear capture. This component is interpreted  as a consequence of Gamov state formation, with the hyperon trapped in the Coulomb field of the residual nucleus.
  Description of such states and their participation in  the capture reaction is presented. Some consequences
are indicated.
\end{abstract}

\PACS{13.75.-n, 25.80.-e, 25.40.Ve}


\section{Introduction}
\label{intro}
An enhancement of events at low $\Sigma^+$ energies, close to the $\Sigma^+$ formation threshold,
was observed by FINUDA \cite{FINU}, in $\Sigma^+ \pi^-$ correlated pairs produced by K$^-$ hadronic captures at-rest on ${}^{6}$Li target. Such phenomenon is absent in the $\Sigma^-$ momentum spectrum of the corresponding $\Sigma^- \pi^+$ sample and this indicates its electromagnetic origin. Monte Carlo simulations of $\Sigma^+$  energy loss in the target, does not seem to properly describe the low $\Sigma^+$ momentum spectrum. A low $\Sigma^+$ momentum peak structure was also measured by AMADEUS (see Ref. \cite{AMAD}) in the reaction:

\begin{equation}
\label{g0}
K^- ~^{12}C ~ \rightarrow  \Sigma^+   ~\pi^-   ~R
\end{equation}
where the residual $R$ is  $ ^{11}$Be when no fragmentation occurs. The low energy ${\Sigma^+}$ events amount to some percent of the total $\Sigma^+ \pi^-$ sample. The solid Carcon fibre target is much thiner in this case, so the $\Sigma^+$ energy loss can not explain the observed phenomenon. Moreover the low momentum structure is not observed in \cite{AMAD} when the K$^-$ is absorbed on a solid ${}^{9}$Be target.

In agreement with old measurements (see review \cite{EMULS}) the spectrum of $\Sigma^+$ momentum
is characterised by a broad distribution centered at $\sim 200$ MeV/c with an additional
threshold enhancement  in the $20\pm15$  MeV/c momentum range, which appears very narrow  on the nuclear momentum scale.
In our analysis, the low momentum peak is  attributed to the interaction of  $\Sigma^+$   with  the residual nucleus. A fraction of the hyperons is trapped  into a Gamov state formed by the interplay among an   attractive nuclear potential and
 the   repulsive Coulomb barrier.
  In section 2 the properties of such states are described, we show that the Gamov state formation offers an explanation to the measured low momentum enhancement in the $\Sigma^+$ distributions.


\section{ The origin of the anomalous threshold peak in the $\Sigma^+$ momentum}
The formation mechanism of the low momentum peak can be described in terms of a sequence of processes:

\begin{itemize}
\item first the $K^-$ meson undergoes hadronic capture   $K^- p \rightarrow  \Sigma^+   ~\pi^- $, which is
 described by a  transition matrix $T$. The residual $R$ is considered as a spectator,

\item the $\Sigma^+ $  is  trapped by the Coulomb potential of  $R$   into a  Gamov state,

\item the Gamov state decays into  $R$ and   $\Sigma^+ $ of low total momentum $\textbf{q}_{\Sigma R}$.
\end{itemize}
The three step mechanism indicated above  is analysed below  in a quasi three-body system consisting of:  meson $K (\pi)$  baryon $p(\Sigma)$ and residual system $R$. We describe the process in the $K^- \,-\, {}^{12}C$ centre of mass system, the initial ${}^{12}C$ nucleus is understood as a bound state of $R$ an $p$.

The following system of Jacobi coordinates will be used:

\begin{itemize}
\item $\textbf{r}$ - which is the relative  $R \,- \,$baryon coordinate,
\item $\textbf{R}$ - which is the relative coordinate of the meson respect to the $R \,- \,$baryon centre of mass.
\end{itemize}
Coordinates referring to the final system are marked with primes.

The final  state is specified by  three  momenta.  The Gamov state system, consisting of the $ R \, -\, \Sigma $ will be denoted by $G$, the total momentum is then
 $ \textbf{p}_G=\textbf{ p}_R + \textbf{p}_{\Sigma}$ and the relative momentum is:
\begin{equation}
\label{s2}
\textbf{q}_{{\Sigma}R} =\alpha'  \textbf{ p}_{\Sigma} ~ - ~\beta'\textbf{ p}_{R}
\end{equation}
where $ \alpha'= m_{R} / (m_{\Sigma}+m_{R})$  and    $ \beta' =m_{\Sigma} / (m_{\Sigma}+m_{R})$.
The third momentum which is necessary to determine the  state is the pion momentum $\textbf{p}_{\pi}$.
The final wave function is  expressed as
\begin{equation}
\label{q5}
\Psi_F =  \exp[i \textbf{R}'(\textbf{p}_{\pi}+ \textbf{p}_{G}-\textbf{p}_K )]\int d\textbf{R}~d\textbf{r} ~G(\textbf{r}',\textbf{r} )~T~\delta(\textbf{r}_{Kp})~ \Phi_K(\textbf{R},\textbf{r} )~\Phi_N(\textbf{r})
\end{equation}
where $G$ is the Green's function describing  propagation of the $\Sigma R$ pair.
For in flight $K^-$ captures the total c.m. system is not fixed, $\textbf{p}_K $ represents  $K^-$ meson momentum in centre of mass system  and
$\Phi_K = \exp(i\alpha \textbf{r}\textbf{p}_K )$.  For atomic captures  $ \textbf{p}_K =0 $ and $\Phi_K$ becomes an atomic function $\Phi_{l}(\textbf{R})$
 for a given angular momentum $l$ state.  The operator $T$   describes  transfer of strangeness. It
is assumed to be of  zero range, as the $KN$ force range is known to be very short,
and depends on the invariant mass
of the meson-baryon pair measured in terms of the final momenta.  Thus $T \equiv  T_{Kp{\Sigma}{\pi}}(M_{\Sigma\pi})$ and the reaction in question offers a chance to study this  energy dependence.
The energy in the $KN$ centre of mass system is  a sum of
the bound nucleon and Kaon energies reduced by  recoil of the pair with respect to the residual system $R.$
The upper kinematic limit of the $M_{\Sigma \pi}$ spectrum is given by $M_N -B_N +E_K$, where $B_N$ in the last nucleon binding energy, $E_K$ is the kaon kinetic energy. For atomic captures the kinematic limit is 1416 MeV in Carbon, for the in-flight capture the corresponding kinematic limit is pushed up of about 14 MeV for $p_K=$120 MeV (tipical momenta of the charged kaons produced at the DA$\Phi$NE factory). This covers the profile  of $\Lambda(1405)$ resonance which dominates the $K^- p$ interaction.

The  Green's function $G$ is   built in terms  of two solutions, regular $\Phi$ and outgoing $\Phi^+$, of the  Schr\"odinger equation
 involving   $\Sigma \,-\, R$  interaction potential.  We split this potential into long   and short ranged parts:
\begin{equation}
\label{s4}
V  =  V_{l}  +i W_{s}   = V_{cul} + V_o \rho (r) +i W_o \rho (r).
\end{equation}
and solve for $G$ in the standard way of the \textit{two potentials problem}.
The short ranged imaginary $i W_{s}$ part describes nuclear absorption of the hyperon.  The long ranged part $V_l$ is composed of the  Coulomb and the nuclear interactions.
For low $Z$ nuclei and low energy hyperon only $S$-wave solutions matter.
Concerning the potential $V_{l}$ two wave function are found.
The radial function $\phi = \Phi/r$ is obtained by solving the Schr\"odinger equation:
\begin{equation}
\label{s5}
- \frac{1}{2\mu} ~\phi(r)''  +  V_l \phi   = E \phi,
\end{equation}
where in Eq.(\ref{s5}) $\mu$ is the  reduced mass of the $\Sigma \,-\, R$ system, and the boundary condition at origin is $\phi(0)=0,\phi'(0)=1$.
The second   solution   of  Eq.(\ref{s5}) (denoted   $\phi^+$)  fulfils the asymptotic condition of an outgoing Coulomb wave.
In practice we cut $V_{cul}$  at large distance and $\phi^+\sim \exp(iq_{\Sigma R}r).$
The Green function  for S-wave  is given by
\begin{equation}
\label{s6}
G_l(E,r,r')   =  \frac{2\mu~[\phi^+(r_{>}) \phi(r_{<})]}{4\pi~[rr' W_l(\phi^+,\phi) ]}
\end{equation}
where  $W_l(\phi^+,\phi)$ is the Wronski determinant, $r_{>}(r_{<})$  denote larger(smaller) of the two cordinates.
The no-interaction  limit (or high momentum limit)   of $ \Phi^+$   is
the spherical Hankel function and similar limit for $\Phi$ is the Bessel function.
Calculations are performed with a  Coulomb potential of uniformly charged sphere,   cut
(at 30 \emph{fm}). The outgoing solution is normalised  at $r_\infty= 50 \emph{fm}$  to its  asymptotic form.

For attractive $V_l$,  localized solutions  may exist at negative energies. For positive energies there  might  exists quasi-localized states due to
the Coulomb barrier i.e. Gamov states.
Such solutions are  characterised by  minute, damped by many orders,  waves outside the barrier and happen at discreet momenta $Q_G$ and energies in a narrow region centered at values  $E_G = (Q_G)^2 /(2\mu) $. The discrete values of momenta characterise the properties
 of the observed  peaks better  than the energies $E_G$,   which are more directly related to resonances that would be observed  in $\Sigma \,-\, R$  scattering
 states.
Zeroes of  the Wronski determinant in the complex energy plane   describe the position of  Green's function singularities
at Gamov  levels $ E_G-i\Gamma_G/2. $
For the low $Z$ systems  in question  well localized solutions may exist.
If the spacial densities $\rho =|\Phi_G(r)/\Phi(0)|^2  $  in such state are  cut  at  $r_\infty $  one can define r.m.s. radii  of the Gamov system.
In addition,  if one requires  $ \rho(r_\infty) < 10^{-6}$ one obtains radii less than $ 10$ fm and  Gamov   levels   in the ($ 0 <  E_G < 0.4 $)  MeV  range.
These states are  coupled  by $T$ to the initial $K$ meson capture states.
Their  widths are very small ($ 0.5 <  \Gamma_G < 20 $) KeV.
The region of $E_G$ indicated above sets  the limits  for the depth of  $V_o$ potential well ($-19.3< V_o < -18.0$) MeV .
In light nuclei  Gamov states  may exist provided there are no bound states. As no $\Sigma$ hyper-nuclear states have been found in the nuclei of interest,
 the $\Sigma$-nuclear potential  is not under  theoretical control.
The experimental investigation of $K^-$ induced reactions in nuclear matter will furnish the real  Gamov state energy and real $V_l$.

The Gamov widths are very narrow, in the KeV region,  while the experimental widths are  about  0.2 MeV.  A natural question arises if 
 the experimental widths are related to the imaginary part of the nuclear potential
which describes the $\Sigma \rightarrow \Lambda$ conversion.
The $W_s $  contributes to the full Green's function  $G$  given by the "two potential"  integral equation which reads in operator form
\begin{equation}
\label{s7}
G = G_l + G_l~ iW_s ~,~ G =  G_l~ [ 1+ iW_s ~G].
\end{equation}
At this stage one has to realize the presence of third body,  the $\pi$ meson.  If one tries to solve eq.(\ref{s7}) in the three body context, say by iterations,  one finds
a propagator projecting on the Gamov state  $ |\Phi><\Phi|/ [E-E_G +i\Gamma_G/2- E^{\pi}(q_\pi)] $.  An integration over intermediate pion momenta
smears the Gamov  singularity and results in a small effect due to small overlap of the Gamov state with the nucleus. Thus the "near singularity" close to the real energy axis  matter in  eq.(\ref{s7})  only  in the Green's  function $G_l$.  We use this approximation  and determine
 $G $, and the  factor  $[ 1+ iW_s ~G]$  from the Schr\"odinger equation (\ref{s5}) solved with the full potential $V$. Such equation also offers discreet  quasi-localized solutions, but only in the far non-physical region for $E'- i\Gamma'/ 2$, where the widths are large $ \Gamma'\sim W_o $, that is in a few  MeV range.  Thus  the  main effect of absorption is the $ i W_sG $ term which renormalizes the strength of coupling to the Gamov state.
 With potential depth $W_s(0) \simeq -15 $ MeV,  characteristic of hyperonic atoms \cite{batt}, one finds $ <|  1+ iW_s ~G|>^2 \simeq 0.6$. This number  describes the loss of $\Sigma$ due to conversion.
  We conclude  that the widths of experimentally observed states are
  not given by nuclear absorption and do not test  the potential $W_s$ directly.
   These widths are due to the $\pi$ meson emission  and follow the distribution
of $ |\textbf{p}_{\Sigma } = \textbf{Q}/\alpha  +\textbf{ p}_{\pi} \beta/\alpha|$  given by eq.(\ref{s2}).  For low momentum hyperons  the pion momentum is almost constant  ($\sim170$  MeV/c) and the width of the
peak is determined by the distribution of  the $\textbf{p}_{\pi}, \textbf{Q}$ angle allowed by the phase space.
With $ Q\rightarrow 0$ the width of peak reduces to a  non-unmeasurably  small $\Gamma_G$.

\section{ Amplitudes, spectral functions}

The transition amplitude generated by the wave function (\ref{s5}) becomes

\begin{equation}
\label{s8}
A =   \frac{T}{W_l} \int d\textbf{r} ~\Phi(r)\Phi_p(r)\Phi_K(\alpha r) \exp(-i\alpha \textbf{p}_{\pi}\textbf{r})
\end{equation}
and requires  wave function for proton $\Phi_p(r)$ (taken from ref. \cite{neff}) and Kaon $\Phi_K$.
For   captures in flight the latter  is assumed  to be  a plane wave.  For  atomic captures one needs to know the distribution over atomic states.
The atomic transition terminates at $ L=2$
\cite{poth} which apparently is the dominant angular momentum at the capture. The distribution of  main quantum numbers is not known
but it  is not relevant as  the absolute capture rates are not measured.

The next step, the spectrum of hyperon momentum is obtained in  standard way integrating over $ d\rho = d_3p/E(p)$ for each particle
\begin{equation}
\label{s9}
P(p_\Sigma)dp_\Sigma = \int ~d\rho^\pi d\rho^\Sigma d\rho^R \delta(\textbf{p}_i-\textbf{p}_{f})\delta(E_i-E_{f})~| A|^2  .
\end{equation}
The shape of the peak is determined by the Wronski determinant.  An excellent  approximation
$|W(q_{\Sigma R} |^2 \simeq \delta(Q-q_{\Sigma R})~ \omega(Q)$
(whith $\omega$ a normalization constant) helps the integration over the three body phase space.

Figure 1. displays the results obtained for the $K^-$ mesons captured in flight. This case is the easiest as the initial mesonic state is known and the
initial $K-nucleus$  interaction is of moderate strength (it was neglected).  The high momentum spectrum was calculated for several versions
of resonant (Breit-Wigner) transition amplitude $T$ modeled to simulate the $\Lambda(1405)$. The dependence on the position of resonance is  moderate.
The position of the peak observed by AMADEUS \cite{AMAD} is obtained with $Q \simeq 15 $ MeV/c.
For all these amplitudes one finds the low peak of  right magnitude of about $3\%$ of the large peak. The shape of the Gamov peak
cannot be tested because the momentum resolution in this region is of several MeV/c. The distribution has a very peculiar structure determined by the Gamov
singularity in the  Wronski determinant, folded over limitations induced by the corner of the available phase space where the peak is located.

\begin{figure}[htb]
\centerline{\includegraphics[height= 5cm,width=7cm]{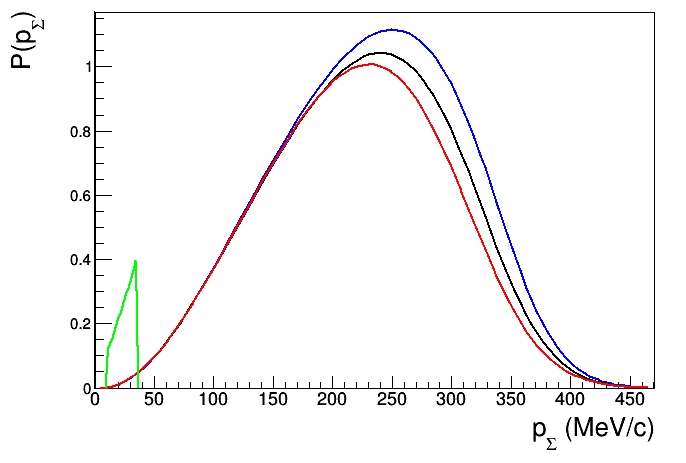}}
\caption{Momentum distribution of $\Sigma^+$ following $K^-$  capture  in-flight on  Carbon nuclei.
The green distribution represents the Gamov peak. Other curves describe the dependence of the spectrum on the position of a 40 MeV wide resonance with $E_r= 1410$ MeV (red distribution), $E_r= 1420$ MeV (black distribution), $T\equiv 1$ (blue distributions).
Normalisations are arbitrary, as only profiles are measured.}
\label{f2}
\end{figure}

\section{Conclusions}
The description of the anomalous low energy momentum distribution  $P(p_{\Sigma})$  in terms of the  Gamov state gives consistent
description of the FINUDA and AMADEUS data.
The mere fact of  existence of the low energy  peak  puts strong limitation on the hyperon nucleus potential.
The peak strength  is more complicated to analyse as it involves : peak position,  $KN$ amplitude and nuclear absorption of
$\Sigma$ hyperons.

These two pieces of information  are, to some extent, a substitute to (apparently nonexistent) $\Sigma$ hypernuclei.
The FINUDA data are related to  hyperon $^5$He system. The only hypernucleus discovered is $^4$He$_\Sigma$, \cite{nagae}. It would
be very  interesting to check whether the Gamov states of $\Sigma^+$ can also be formed in heavier nuclei.
Positive answer would open a new branch of hypernuclear spectroscopy.

The full understanding of  $p_{\Sigma}$ spectrum, in particular the weight  of the anomalous peak, relative to the standard spectrum,
might be more informative than the $\Sigma$ hypernuclear spectroscopy would be.  New experiments would be very helpful.



\noindent \emph{Acknowledgements} We thank Nicolas Keeley  for useful advice and Stefano Piano for consultations.

\end{document}